\documentstyle[aps,epsfig]{revtex}
\newcommand{\hc}{{\mathrm h.c.}}
\newcommand{\tr}{{\mathrm tr}}
\newcommand{\re}{{\mathrm Re}}
\newcommand{\im}{{\mathrm Im}}

\begin{document}
\title{Unitarity triangles and geometrical description of CP violation with
Majorana neutrinos}
\author{J. A. Aguilar-Saavedra} 
\address{Departamento de F\'{\i}sica Te\'{o}rica y del Cosmos,
Universidad de Granada,
E-18071 Granada, Spain}
\author{G. C. Branco}
\address{Departamento de F\'{\i}sica and CFIF,
Instituto Superior T\'ecnico,
1096 Lisboa, Portugal}
\date{\today}
\maketitle
\begin{abstract}
We generalize the geometrical description of CP violation in the Standard Model
in terms of a unitarity triangle. For three left-handed Majorana
neutrinos CP violation in the lepton sector is determined by three unitarity
triangles. With three additional right-handed neutrinos 
15 quadrangles are required to characterize CP violation. We show the
relation of the unitarity polygons with physical observables.

\draft
\pacs{11.30.Er, 12.60.-i, 14.60.Pq, 14.60.St}

\end{abstract}
\section{Introduction}
The recent Superkamiokande data \cite{papiro1} on atmospheric neutrinos provides
evidence for neutrino oscillations, thus suggesting that neutrinos have
non-vanishing masses. The simplest way of understanding the smallness of these
masses is through the seesaw mechanism \cite{papiro5} which naturally leads to
Majorana neutrinos.

In this article we provide a geometrical interpretation of CP violation with
Majorana neutrinos. In the quark sector of the three-generation Standard Model
(SM) it is well known that CP violation can be described by six
unitarity triangles \cite{papiro4b}. They are obtained from orthogonality of the
rows and columns of the Cabibbo-Kobayashi-Maskawa matrix $V$ \cite{papiro2},
and are all equivalent.
Under a rephasing transformation of the quark fields, these unitarity triangles
rotate and therefore their orientation has no physical meaning. 
However, their area does have physical meaning and in fact all the six triangles
have the same area which is proportional to the rephasing invariant CP violating
quantity $|\im \, V_{ij} V_{kj}^* V_{kl} V_{il}^*|$ \cite{papiro3}. It is natural
to ask how this geometrical analysis can be extended to the leptonic sector, when
Majorana neutrinos are present. In this article we will address to this question,
considering first the cases of three and four left-handed Majorana neutrinos and
then the case of three left-handed and an arbitrary number of right-handed
neutrinos. We will interpretate the well-known features of CP violation for
Majorana neutrinos in terms of geometrical properties of polygons.

\section{Geometrical interpretation for three left-handed neutrinos}

In the SM extension with Majorana neutrino masses, the mass terms
of the leptonic Lagrangian, in the weak eigenstate basis, can be written as
\begin{equation}
-{\mathcal L}_{\mathrm mass} = \bar l_L^0 M_l l_R^0 + \frac{1}{2} \bar \nu_L^0
M_L (\nu_L^0)^c + \hc \,,
\label{ec:2}
\end{equation}
where $l_L^0$, $l_R^0$ and $\nu_L^0$ are three-dimensional vectors in flavour
space, $M_l$ and $M_L$ are $3 \times 3$ complex matrices and $(\nu_L^0)^c = C
(\bar \nu_L^0)^T$. $M_L$ can be taken as symmetric without loss of generality.
The mass term for the neutrinos can be generated extending the scalar sector
with a Higgs triplet or can be the result of a seesaw mechanism
\cite{papiro5}. The charged current term in this basis is
\begin{equation}
-{\mathcal L}_{\mathrm CC} = \frac{g}{\sqrt 2} \bar \nu_L^0 \gamma_\mu l_L^0
W_\mu^\dagger + \hc
\label{ec:3}
\end{equation}
The leptonic mass matrices $M_l$, $M_L$ are diagonalized through the
transformations $l_L^0 = U_L^l l_L$, $\nu_L^0 = U_L^\nu \nu_L$,
$l_R^0 = U_R^l l_R$, with $U_L^{l,\nu}$, $U_R^l$ $3 \times 3$ unitary matrices,
so that $U_L^{l \dagger} M_l U_R^l = D_l$,
$U_L^{\nu \dagger} M_L U_L^{\nu *} = D_L$ where $D_l$, $D_L$ are the diagonal
mass matrices. One can then write, in the mass eigenstate basis,
\begin{eqnarray}
-{\mathcal L}_{\mathrm mass} & = & \bar l_L D_l l_R + \frac{1}{2} \bar \nu_L
D_L (\nu_L)^c + \hc \,, \nonumber \\
-{\mathcal L}_{\mathrm CC} & = & \frac{g}{\sqrt 2} \bar \nu_L \gamma_\mu U l_L
W_\mu^\dagger + \hc 
\label{ec:4}
\end{eqnarray}
The mixing matrix $U=U_L^{\nu \dagger} U_L^l$
is the analogous of the CKM matrix in the quark
sector, and can be parametrized with 3 mixing angles and 6 phases. The phase
of the charged lepton mass eigenstates is arbitrary, and we can use this freedom
to eliminate three of these phases. However, 
the phase of the neutrino mass eigenstates cannot be changed, since this
transformation does not leave invariant the neutrino
mass matrix in Eq. (\ref{ec:4}). Hence the
mixing matrix $U$ has in general three independent, CP violating
physical phases instead of
one. Two of these phases are ``Majorana phases'' that could be removed if
rephasing of the neutrino fields was allowed. The other phase is the analogous
to the CKM phase which cannot be removed even for Dirac neutrinos, a ``Dirac
phase''. 

The mixing matrix
can be parametrized in general as a diagonal matrix of phases
multiplied by a 
$3 \times 3$ unitary matrix in the standard parametrization \cite{papiro4}
\widetext
\begin{equation}
U = \left( \begin{array}{ccc}
1 & 0 & 0 \\
0 & e^{i \alpha_2} & 0 \\
0 & 0 & e^{i \alpha_3}
\end{array} \right)
\cdot \left( \begin{array}{ccc}
c_{12} c_{13} & s_{12} c_{13} & s_{13} e^{-i \delta} \\
-s_{12} c_{23} -c_{12} s_{23} s_{13} e^{i \delta} & 
c_{12} c_{23} -s_{12} s_{23} s_{13} e^{i \delta} & s_{23} c_{13} \\
s_{12} s_{23} -c_{12} c_{23} s_{13} e^{i \delta} &
-c_{12} s_{23} -s_{12} c_{23} s_{13} e^{i \delta} & c_{23} c_{13}
\end{array} \right) \,.
\label{ec:5}
\end{equation}
In this parametrization, $\delta$ is the Dirac phase and $\alpha_2$, $\alpha_3$
are the Majorana phases. CP is conserved if $\delta = 0$ mod $\pi$ and
$\alpha_2=0$,
$\alpha_3=0$ mod $\pi/2$ (from now on we will write for simplicity $\delta = 0$,
$\alpha_2 = 0$, $\alpha_3=0$). Note that Majorana phases of $\pm \pi/2$ do not
imply CP violation but indicate different CP parities of the neutrino mass
eigenstates \cite{papiro10}.

The only rephasing transformations allowed are 
$l_{Lj,Rj} \to e^{i \lambda_j} l_{Lj,Rj}$
Under these transformations, the matrix elements of $U$ 
transform as $U_{ij} \to e^{i \lambda_j} U_{ij}$.
Hence the minimal rephasing invariant terms are the products
$U_{ij} U_{kj}^*$, and the minimal CP violating
quantities their imaginary parts $\im \, U_{ij} U_{kj}^*$
\footnote{Strictly speaking, $\im \, U_{ij} U_{kj}^*$ are not CP violating if
$\re \, U_{ij} U_{kj}^* = 0$ (see Section 4).}.

One can define triangles analogous to those depicted in the quark sector by
multiplying two columns of $U$, {\em e. g.}
$U_{11} U_{13}^* + U_{21} U_{23}^* + U_{31} U_{33}^* = 0$. 
Under rephasing transformations, these triangles rotate in the complex plane,
$U_{ij} U_{ik}^* \to e^{i (\lambda_j-\lambda_k)} \; U_{ij} U_{ik}^*$, so their
orientation has no physical meaning. They share a common area 
$A = 1/2 \; |\im \, U_{ij} U_{kj}^* U_{kl} U_{il}^*|$. We will call these
triangles ``Dirac triangles''. The vanishing of their
area implies ${\mathcal J}_U \equiv |\im \, U_{ij} U_{kj}^* U_{kl} U_{il}^*|= 0$,
but does not imply that the minimal CP violating quantities
$\im \, U_{ij} U_{kj}^*$ are zero, and CP can still be violated. In terms of 
phases, ${\mathcal J}_U=0$ implies that the Dirac phase vanishes
($\delta = 0$ in the parametrization in Eq. (\ref{ec:5})) but the Majorana
phases can still violate CP. Thus these
triangles provide a necessary but not sufficient condition for CP conservation
and are not enough to completely describe CP violation.

\begin{figure}[htb]
\begin{center}
\mbox{\epsfig{file=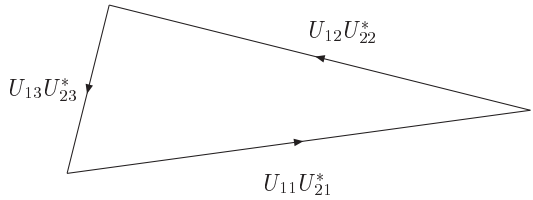,width=7cm}}
\end{center}
\caption{Majorana unitarity triangle $T_{12}$. Its orientation is fixed by the
Majorana phases and it cannot be rotated in the complex plane.
\label{fig:2} }
\end{figure}

One can also define three ``Majorana triangles'' multiplying two rows of $U$
(see Fig. \ref{fig:2}):
\begin{eqnarray}
T_{12} & \equiv & U_{11} U_{21}^* + U_{12} U_{22}^* + U_{13} U_{23}^* = 0 \,,
\nonumber \\
T_{13} & \equiv & U_{11} U_{31}^* + U_{12} U_{32}^* + U_{13} U_{33}^* = 0 \,,
\nonumber \\
T_{23} & \equiv & U_{21} U_{31}^* + U_{22} U_{32}^* + U_{23} U_{33}^* = 0 \,.
\end{eqnarray}

Under a change of phase these triangles do not rotate in the complex plane,
since all their terms are rephasing invariant. Thus their orientation is
physically meaningful. These Majorana triangles provide
the necessary and sufficient conditions for CP conservation:
\begin{enumerate}
\item Vanishing of their common area $A={\mathcal J}_U/2$.
\item \label{cond2} Orientation of all Majorana triangles along the direction of the real or
imaginary axes.
\end{enumerate}
The first condition implies that the three triangles collapse
into lines in the complex plane and the Dirac phase vanishes.
Condition (2) implies that
the Majorana phases do not violate CP.
If the three collapsed triangles are on the $x$ axis, one has
$\im \, U_{ij} U_{kj}^* = 0$ $\forall i,j,k$ and CP is obviously conserved. 
If one of these triangles, $T_{ik}$, is parallel to the $y$ axis,
that means that the
mass eigenstates $\nu_i$ and $\nu_k$ have opposite CP parities, but there is not
CP violation. Multiplying the $i$ or the $k$ row
by $\pm i$ we can rotate the triangle to the $x$ axis making the mass of the
corresponding eigenstate negative. Hence the three Majorana triangles
provide
a complete description of CP violation. Each one of their sides, if not parallel
to one of the axes, is itself a signal of CP violation,
contrarily to the Dirac case where only a nonzero area signals CP violation.

One may wonder whether the three conditions (\ref{cond2}) are independent,
{\em i. e.}
whether the orientation of the three triangles is related. If at least one
column of $U$, {\em e. g.} the first column, has nonzero elements, one has
\begin{equation}
\arg U_{11} U_{31}^* =  \arg U_{11} U_{21}^* + \arg U_{21} U_{31}^* ~~
\mathrm{mod}~ 2\pi \,.
\label{ec:7}
\end{equation}
This is true independently of the area $A$, but
in the case of vanishing areas, Eq. (\ref{ec:7}) implies that the angles between
the triangles and the real axis
are related, and only two triangles are needed, for
instance $T_{12}$ and $T_{13}$.
Requiring that these two triangles are on the $x$ or $y$ axis eliminates the two
Majorana phases.

To describe CP violation in the most general cases,
the three triangles are needed. In order to see that this is the case, let us
consider the mixing matrix
\begin{equation}
U = \left( \begin{array}{ccc}
1 & 0 & 0 \\
0 & 1/\sqrt 2 & 1/\sqrt 2 \\
0 & e^{i \alpha}/\sqrt 2 & -e^{i \alpha}/\sqrt 2
\end{array} \right) \,.
\label{ec:8}
\end{equation}
For this particular matrix, $T_{12}$ and $T_{13}$ are trivial,
and CP is violated unless
$\alpha = 0$ mod $\pi/2$, {\em i. e.} $T_{23}$ is parallel to one of the axes.
Analogous mixing matrices can be written to show that $T_{12}$ and $T_{13}$ are
necessary in general.

To conclude this Section, let us discuss two interesting special cases. If
one neutrino is massless, {\em e. g.} $m_{\nu_1} = 0$,
the phase of $\nu_1$ can be changed leaving
${\mathcal L}_{\mathrm mass}$ invariant. The orientation of the
triangles
$T_{12}$ and $T_{13}$ has no physical meaning, and all CP violation can be
summarized in $T_{23}$. In this case we have only the Dirac phase ($\delta$) and
one Majorana phase ($\alpha_3$).
Requiring that the area of $T_{23}$ is zero eliminates
$\delta$, and requiring that $T_{23}$ is parallel to
the $x$ or $y$ axis eliminates
$\alpha_3$. The cases $m_{\nu_2} = 0$ and $m_{\nu_3} = 0$ are similar.

In the case that there is one and only one zero in $U$, {\em e. g.} $U_{31}$,
the three triangles have null area and the Dirac phase is zero.
The orientation of the
three triangles is related and only two of them are necessary to describe CP
violation. If there are two zeroes in the mixing matrix, $U$ is analogous
to the matrix in Eq. (\ref{ec:8}) with
at most one Majorana phase, and only one triangle is nontrivial.

\section{Generalizations}

\subsection{Four left-handed neutrinos}

Before considering the inclusion of right-handed neutrinos it is convenient to
analyze the simpler case of four left-handed neutrinos.
The CKM matrix in this case is a $4 \times 4$ matrix, with three Dirac phases
and three Majorana phases. The unitarity
relations between its rows can be represented in the complex plane as
six Majorana quadrangles. For example, the orthogonality condition between the
first and second rows can be represented as the quadrangle
\begin{equation}
Q_{12} \equiv U_{11} U_{21}^* + U_{12} U_{22}^* + U_{13} U_{23}^* 
+ U_{14} U_{24}^* = 0 \,.
\label{ec:9}
\end{equation}
Its area is
\footnote{For simplicity we assume that $Q_{12}$ when drawn in the order of Eq.
(\ref{ec:9}) is convex. In other case, its sides must be reordered, obtaining an
expression similar to Eq. (\ref{ec:10}). The results are unchanged.}
\begin{eqnarray}
A_{12} & = & \frac{1}{4} \{ 
|\im \, U_{11} U_{21}^* U_{22} U_{12}^*| + 
|\im \, U_{12} U_{22}^* U_{23} U_{13}^*| \nonumber \\
& & + |\im \, U_{13} U_{23}^* U_{24} U_{14}^*| 
+ |\im \, U_{14} U_{24}^* U_{21} U_{11}^*| \} \,.
\label{ec:10}
\end{eqnarray}
If $A_{12} = 0$, the four imaginary products are zero. The condition for the
vanishing of the three Dirac phases is that the areas of three independent
convex
quadrangles (for instance, $Q_{12}$, $Q_{23}$, $Q_{34}$ or $Q_{12}$, $Q_{13}$,
$Q_{23}$) are zero \cite{papiro6}. When this condition is fulfilled, the areas
of all the quadrangles are zero. To describe CP violation with all generality
the six quadrangles $Q_{12}$, $Q_{13}$, $Q_{14}$, $Q_{23}$, $Q_{24}$, $Q_{34}$
are necessary, since there are special
cases when all quadrangles but one are trivial.
CP is conserved if, and only if, these six convex quadrangles have null area and
are orientated in the direction of one of the axes. However, if at least one
column of $U$ has all elements non-vanishing, {\em e. g.} the first column,
\widetext
\begin{eqnarray}
\arg U_{11} U_{31}^* & = &  \arg U_{11} U_{21}^* + \arg U_{21} U_{31}^* ~~
\mathrm{mod} ~ 2\pi \,, \nonumber \\
\arg U_{21} U_{41}^* & = &  \arg U_{21} U_{31}^* + \arg U_{31} U_{41}^* ~~
\mathrm{mod} ~ 2\pi \,, \nonumber \\
\arg U_{11} U_{41}^* & = &
\arg U_{11} U_{21}^* + \arg U_{21} U_{31}^* + \arg U_{31} U_{41}^*
~~ \mathrm{mod} ~ 2\pi \,,
\end{eqnarray}
and the three quadrangles $Q_{12}$, $Q_{23}$ and $Q_{34}$ completely
characterize CP violation.

\subsection{Three left-handed and $n_R$ right-handed neutrinos}

\begin{figure}[htb]
\begin{center}
\mbox{\epsfig{file=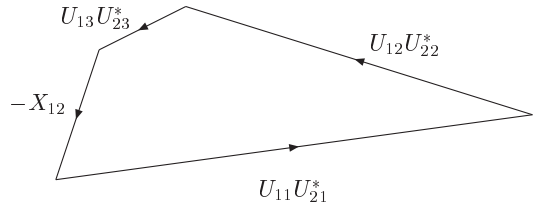,width=7cm}}
\end{center}
\caption{Majorana unitarity quadrangle $Q_{12}$ for an arbitrary number of
right-handed neutrinos $n_R$. Its orientation is fixed by the
Majorana phases and it cannot be rotated in the complex plane.
\label{fig:3}}
\end{figure}

The case of 3 left-handed and $n_R$ right-handed neutrinos is similar to the
extension of the SM with $n_R$ up-type quark singlets, with some differences
due to the Majorana character of the neutrinos. The CKM mixing matrix $U$ is a
$(3+n_R) \times 3$ submatrix of a $(3+n_R) \times (3+n_R)$ unitary matrix, with
$2 n_R+1$ Dirac phases and $n_R+2$ Majorana phases. In addition, the neutral
current Lagrangian contains nondiagonal terms,
\begin{equation}
-{\mathcal L}_{\mathrm NC} = \frac{g}{2c_W} \bar \nu_L \gamma^\mu X \nu_L Z_\mu
+ \hc \,,
\end{equation}
where $X$ is a $(3+n_R) \times (3+n_R)$ hermitian matrix
with complex nondiagonal elements. Note that $\nu_L$ is a linear combination of
weak eigenstates $\nu_L^0$, $(\nu_R^0)^c$ with different isospin.
However, the flavour-changing neutral (FCN) couplings do not contain
additional CP violating phases. For any number of
right-handed neutrinos $n_R$, the unitarity relations between rows of $U$ can be
represented as convex Majorana quadrangles in the complex plane \cite{papiro7}.
For example, the orthogonality condition
between the first and second rows reads
\begin{equation}
Q_{12} \equiv U_{11} U_{21}^* + U_{12} U_{22}^* + U_{13} U_{23}^* 
 = X_{12} \,,
\end{equation}
with $X_{12}$ the FCN coupling between the neutrino
mass eigenstates $\nu_1$ and $\nu_2$ (see Fig. \ref{fig:3}).

The geometrical interpretation is then very similar to the previous case. The
condition for the vanishing of the $2n_R+1$ Dirac phases is that the area of
$(n_R + 2)(n_R + 1)/2$ independent convex quadrangles is zero, for instance
$Q_{12}$, $Q_{23}$ \dots $Q_{n_R+2,n_R+3}$. In this case, the areas of the
remaining quadrangles are also zero. The condition for the vanishing of the
$n_R+2$ Majorana phases is that all the $(n_R + 3)(n_R + 2)/2$ quadrangles are
parallel to the $x$ or $y$ axis. Hence, the complete description of CP violation
with three left-handed and $n_R$ right-handed neutrinos is achieved with
$(n_R + 3)(n_R + 2)/2$ quadrangles. Note that in the limit where the heavy
neutrinos decouple, $X_{ij} = \delta_{ij}$ for $i,j=1,2,3$ and the three
quadrangles involving only the light neutrinos reduce to the three triangles
described in the previous Section. The other quadrangles are trivial.

\section{Unitarity polygons and physical observables}

Relevant information on the unitarity polygons (and the CP violating phases) can
be extracted from CP violating and CP conserving processes. The CP violating
processes may be sensitive to the Majorana phases or may not. We will show
examples of each case.

CP violation in neutrino oscillations is not sensitive to Majorana phases. 
The Dirac or Majorana character of the neutrinos is not revealed in this kind of
experiments, and thus CP violation observables are proportional to imaginary
quartets $\im \, U_{ij} U_{kj}^* U_{kl} U_{il}^*$. Neutrino oscillations can
only provide information on the areas of the unitarity polygons and the Dirac
phases.

\begin{figure}[htb]
\begin{center}
\mbox{\epsfig{file=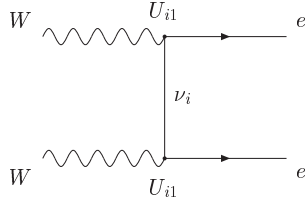,width=4cm}}
\end{center}
\caption{Feynman diagram for neutrinoless double beta decay mediated by Majorana
neutrinos.
\label{fig:5}}
\end{figure}

One important process to test the Majorana character of the neutrinos is
neutrinoless double-beta decay. This process is mediated by the diagram in
Fig. \ref{fig:5}, where we can observe that the Majorana character of the
neutrinos is essential. 

The cross-section of this process is
\begin{equation}
\sigma = C \, \sum_{i,j} m_{\nu_i} m_{\nu_j} (U_{i1} U_{j1}^*)^2 \,,
\label{ec:15}
\end{equation}
where the rephasing invariance is explicit. $C$ is a factor independent of
the mixing angles. There are terms proportional to
$|U_{i1}|^4$ and also terms proportional to the first sides of the triangles (or
quadrangles), the sides involving the couplings of the neutrinos to the
electron. The measurement of the neutrinosless double beta decay rate then
serves to constrain the CP violating parameters.

\begin{figure}[htb]
\widetext
\begin{center}
\raisebox{-1.75cm}{\mbox{\epsfig{file=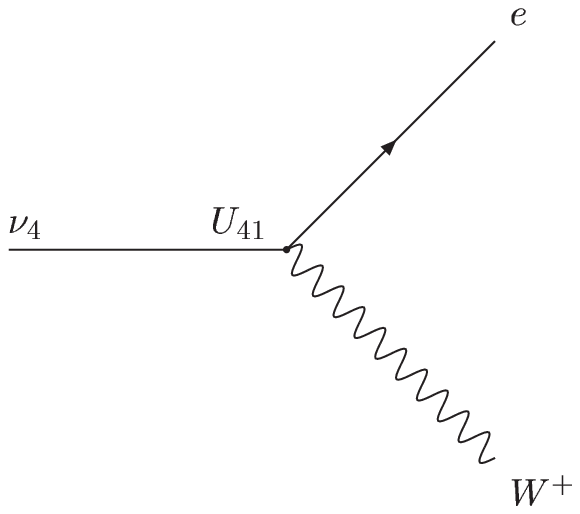,width=4cm}}} ~~+~~~~
\raisebox{-1.75cm}{\mbox{\epsfig{file=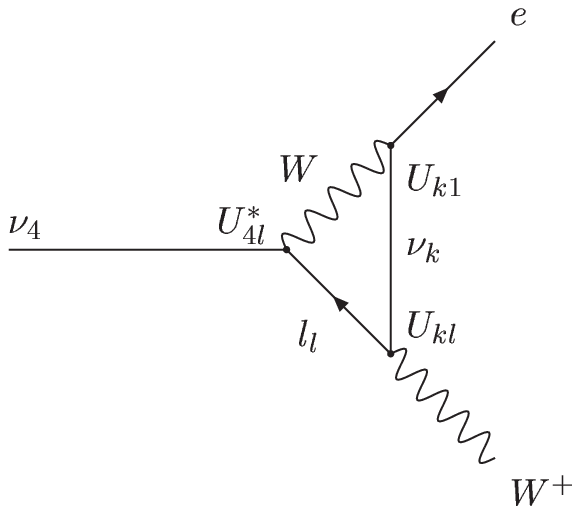,width=4cm}}}
\end{center}
\caption{Tree-level and one-loop contributions to the CP asymmetry in the decay
of a heavy Majorana neutrino $\nu_4 \to W^+ e$.
\label{fig:4}}
\end{figure}

The decay of heavy Majorana neutrinos has been proposed as a source of CP
violation for baryogenesis \cite{papiro8}. Here for definiteness
we will consider the CP
asymmetry in the decay of a heavy neutrino $\nu_4 \to W^+ e$.
The tree-level and one-loop diagrams relevant for the CP asymmetry in this
decay are shown in Fig. \ref{fig:4}. 
Note that the Majorana character of the neutrinos plays an
essential r\^ole in the second diagram.
There are four more one-loop diagrams not taken into account. Two of them
involve $Z$ and photon exchange between the electron and the $W$. They have the
same weak phase as the tree-level diagram and do not contribute to the CP
asymmetry at lowest order. The other two involve $Z$ exchange between the
neutrino and the electron and $W$, respectively. If FCN couplings are neglected,
they also have the same weak phase as the tree-level diagram.

The difference $\Gamma(\nu_4 \to W^+ e)-\bar \Gamma(\nu_4 \to W^- e^+)$ is
a CP violating observable. Extracting the dependence on the mixing angles it
can be written as
\begin{equation}
\Gamma - \bar \Gamma = \sum_{k,l} f_{kl} \, \im \, (U_{41} U_{k1}^*) \,
(U_{4l} U_{kl}^*) \,,
\label{ec:14}
\end{equation}
with $f_{kl}$ form factors independent of mixing angles.
To understand this formula we observe in Fig. \ref{fig:4} that the mixing angle
factors of the tree-level and one-loop contributions, $U_{41}$ and
$U_{kl} U_{4l}^* U_{k1}$ respectively, transform in the same way under
rephasing of the charged lepton fields (as they must), but not under rephasing
of the neutrino mass eigenstates (as they should in the case of Dirac
neutrinos).
If the two mixing angle factors transformed in the same way under rephasing of
the neutrino mass eigenstates, we would have a dependence
$\Gamma - \bar \Gamma \propto \im \, U_{41} U_{k1}^* U_{kl} U_{4l}^*$ or
similar,
and the asymmetry would not be sensitive to the Majorana phases.
The diagrams involving $Z$ exchange between the neutrino and
$e$, $W$ give an additional contribution of this type.

The CP asymmetry in Eq. (\ref{ec:14}) is written as a linear combination of
imaginary parts of products of the sides $1,l$ of the 
quadrangle $Q_{4k}$, summing over $k$ and $l$. The terms with $l=1$ contain
$\im \, (U_{41} U_{k1}^*)^2 = 2 \, \re \, (U_{41} U_{k1}) \, \im \, (U_{41}
U_{k1})$. We see that a side of $Q_{4k}$ does not contribute to the CP
asymmetry if it is parallel to one of the axes, although it may contribute
in the terms of mixing with other sides if the Dirac phases are nonzero,
{\em i. e.} if other sides of the same quadrangle are CP violating.

Finally we will show the relation between the unitarity triangles and the
weak-basis invariants. The lowest order weak-basis invariant for three
left-handed Majorana neutrinos is \cite{papiro9}
\begin{eqnarray}
I & = & \im \, \tr \, M_l M_l^\dagger M_L M_l^* M_l^T M_L^* M_L M_L^*
\nonumber \\
& = & \sum_{i,j,k,l} m_j^2 m_l^2 m_{\nu_i}^3 m_{\nu_k} \im \,
(U_{ij} U_{kj}^*) \, (U_{il} U_{kl}^*) \,,
\label{ec:16}
\end{eqnarray}
with $m_i$ the mass of the charged lepton $i$. In Eq. (\ref{ec:16}) rephasing
invariance is explicit. The invariant is written as 
a sum of imaginary parts of products of two sides $j,l$ of the same triangle
$T_{ik}$ weighted by mass factors. The terms with $j=l$ are proportional to
$\im \, (U_{ij} U_{kj}^*)^2 = 2 \, \re \, (U_{ij} U_{kj}^*) \, \im \, (U_{ij}
U_{kj}^*)$. We see that a side of $T_{ik}$ parallel to the real or imaginary
axis does not contribute to $I$. If the Dirac phase is also zero, the whole
contribution of $T_{ik}$ is zero. Thus, although the minimal CP violating
quantities are $\im \, U_{ij} U_{kj}^*$, they need to interfere
with their real part or with other sides of unitarity triangles.

\acknowledgements
This work was partially supported by CICYT under contract AEN96--1672 and by the
Junta de Andaluc\'{\i}a, FQM101.

\end{document}